# GANIL STATUS REPORT

B. Jacquot , F. Chautard, A.Savalle, & Ganil Staff
GANIL-DSM/CEA,IN2P3/CNRS, BP 55027, 14076 Caen Cedex, France

*Abstract*

The GANIL-Spiral facility (Caen, France) is dedicated to the acceleration of heavy ion beams for nuclear physics, atomic physics, radiobiology and material irradiation. The production of radioactive ion beams for nuclear physics studies represents the main part of the activity. The facility possesses a versatile combination of equipments, which permits to produce accelerated radioactive ion beams with two complementary methods: Isotope Separation In Line (ISOL) and In-Flight Separation techniques (IFS). Considering the future of GANIL, SPIRAL II projects aims to produce high intensity secondary beams, by fission induced with a 5 mA deuteron beam on an uranium target.

## INTRODUCTION

At present, the GANIL facility produces a wide spectrum of high intensity ion beams ranging from $^{12}$C to $^{238}$U accelerated up to 96 MeV/A. The acceleration scheme lies on the use of three cyclotrons in line (C01 or C02 , CSS1 ,CSS2 ). The large number of experimental set-ups in the experimental hall allows to cover a large scientific spectrum. GANIL has a long-standing experience to produce radioactive ion beams by in-flight fragmentation of a high energy beam. With this technique, GANIL delivers radioactive ion beams between 40 MeV/A and 90 MeV/A. Since September 2001, SPIRAL produces in a thick target radioactive atoms which are ionized by an ECR source. The low energy radioactive beams are subsequently post-accelerated by the cyclotron CIME in a lower energy domain : [1.7, 25] MeV/A.

The operation and the running statistics of GANIL-SPIRAL are presented, with particular attention to the first SPIRAL beams. Few results about the cyclotron CIME, as the mass selection and tuning principle are summarized.

We also present recent development and modification of the facility which aim to increase beam intensity or beam time for physics experiments.

## OPERATION STATISTICS

The GANIL accelerators are operated 34 to 35 weeks (5500 hours) per year. The operation statistics, over these last three years, are given bellow.

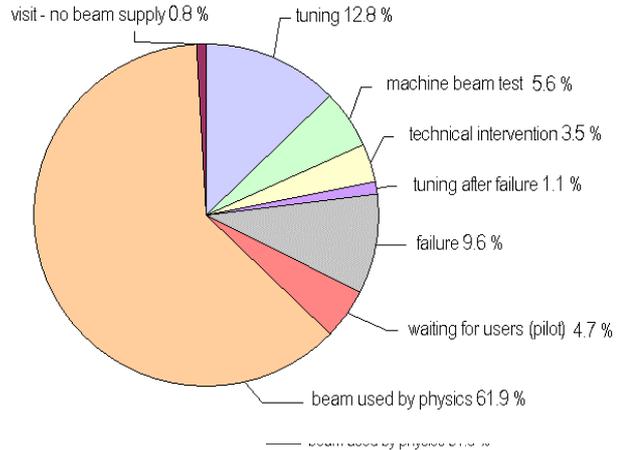

Figure 1: Operation statistics in years 2001-2003

Despite a greater number of beam changes and beam studies, due to operation of CIME cyclotron, the time available for physics experiments is slightly increasing. Indeed, beam tuning or machine studies with CIME cyclotron are scheduled in parallel of experiments with the other cyclotrons. The inverse is also possible. The available beam time is shared between different user categories : nuclear physics (85% of available beam time), atomic physics, radiobiology and material irradiation. In addition to experiments using high energy beams, the Intermediate Energy Exit (SME, Fig.2), working simultaneously with the main beam, uses about 1400 beam hours per year for atomic physics.

A new beam line (IRRSUD, fig 2.), dedicated to material irradiation and radiobiology, is operational since November 2002. This irradiation facility uses the 1MeV/A beam coming from one of the compact cyclotrons while the second compact cyclotron injects a beam in the separated sector cyclotrons.

Therefore the GANIL facility is able to deliver three simultaneous stable beams, at low, medium and high energy for three experiments in parallel.

The main part of the activity concerns production of radioactive ion beams for nuclear physics studies.

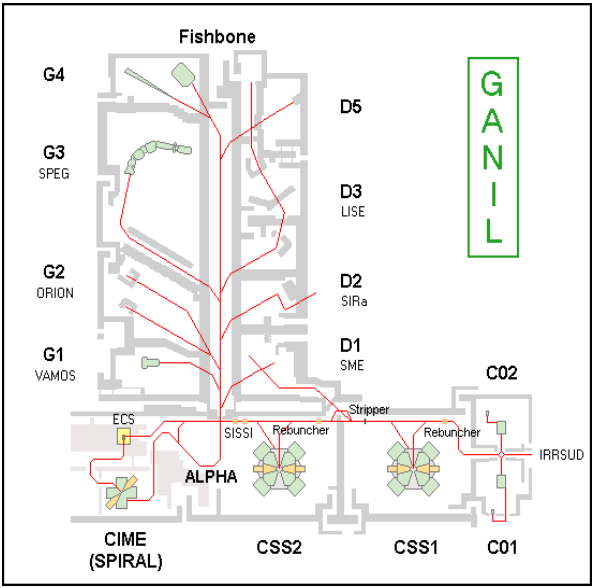

Figure 2 : Map of GANIL showing accelerators and experimental areas. C01 and C02 are compact cyclotrons (E<1 MeV.A); CSS1 (E<13.7 MeV.A) and CSS2 (E<96 MeV.A) are separated sectors cyclotrons (K=380).

## RADIOACTIVE ION BEAM

### IN-FLIGHT FRAGMENTATION

For creating high energy radioactive ion beam by in-flight projectile fragmentation two rotating targets and associated separators are available. The first target, SISSI, is sandwiched by a set of 2 super-conducting solenoids [1]. The strong focalisation induced by the 11 Tesla field minimized the emittance increase in the target. The produced fragments can be selected and purified with a the Alpha spectrometer. The beam quality is given by the acceptance of the beam line (The transverse emittance is limited to 16π.mm.mrad , while the momentum dispersion Δp/p is limited to ±0.5% ).

A second rotating target is available upstream the LISE[2] spectrometer. This target is being up-graded to increase the maximal intensity of the incoming beam. The associated fragment separator LISE have better momentum acceptance (Δp/p=2.5%). The maximal magnetic rigidity of LISE was quite limitative for light fragments, therefore a new branch of the spectrometer, called LISE2000, has been designed to reach 4.3 Tm.

### SPIRAL BEAMS [3]

SPIRAL facility produces radioactive ion beams with the so called ISOL techniques : The stable heavy ion beams of GANIL are sent onto a target and source assembly. The radioactive atoms produced by nuclear reactions are released from the target, kept at high temperature, into an ECR source. After ionization and extraction from the source (extraction voltage < 34 kV), the multi-charged radioactive ions are accelerated up to a maximum energy of 25 MeV/A by the compact cyclotron CIME (K=265). The list of radioactive beams produced and accelerated with SPIRAL from September 2001 , until October 2004 are given in table 1. Up to now, the primary beam power on SPIRAL targets has been limited to 1.4 kW for thermal and safety reasons. A new target, designed for 3 kW of 13C, has been successfully tested.

| SPIRAL BEAMS | | |
|---|---|---|
| Ion | Energy (MeV/A) | Intensity (pps) |
| $^{6}He^{1+}$ | 3.2, 5 | $3 \times 10^{+7}$ |
| $^{8}He^{2+}$ | 15.4 | $2 \times 10^{+4}$ |
| $^{8}He^{1+}$ | 3.4 to 3.9 | $8 \times 10^{+4}$ |
| $^{18}Ne^{4+}$ | 7 | $1 \times 10^{+6}$ |
| $^{24}Ne^{5+}$ | 4.7, 10 | $2 \times 10^{+5}$ |
| $^{26}Ne^{5+}$ | 10 | $3 \times 10^{+3}$ |
| $^{74}Kr^{11+}$ | 2.6 | $1.5 \times 10^{+4}$ |
| $^{76}Kr^{11+}$ | 2.6, 4.4 | $6 \times 10^{+5}$ |
| $^{44}Ar^{9+}$ | 10.8 | $2.0 \times 10^{+5}$ |
| $^{46}Ar^{9+}$ | 10.8 | $2.0 \times 10^{+4}$ |

Table 1 : SPIRAL beams until October 2004. Let us note that there is a strong demand on $^{8}He^{2+}$ beam.

Let us mention that a new beam line has been constructed to inject the low energy radioactive beam in a ensemble of RFQ gas cooler followed by a Paul trap.

### CIME cyclotron as post-accelerator

The cyclotron CIME has an axial injection system. Two different injection centres are needed to accelerate ion beams from 1.7 to 25 MeV/A at the RF harmonics[2] H=2,3,4 and 5.

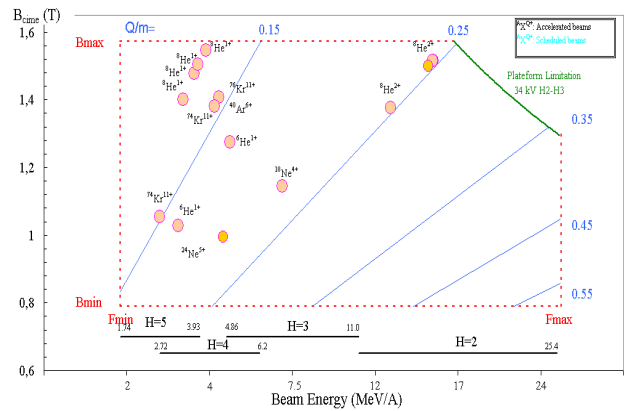

Figure3 : CIME working diagram including accelerated beams (April 2004)

The low-energy beams, from 1.7 to 6.2 MeV/A, are injected at a radius R=45 mm, using a Pabot-Belmont

---

[2] Radio-Frequency Harmonics : ratio between the frequency of the accelerating cavities and the ion revolution frequency in the cyclotron.

inflector complemented by an electrostatic quadrupole, and accelerated on RF harmonics 4 or 5.

The high-energy beams, from 4.9 to 25.4 MeV/A, are injected at a radius R=34 mm, using a Muller inflector and accelerated on RF harmonics 2 or 3.

## CIME cyclotron as separator[4]

As far as the purification of very rare isotope is concerned the achievement of very good mass resolution is not sufficient. Since stable contaminants can dominate by several orders of magnitude the intensity of given radioactive ion species, it is of utmost importance to perform a very clean separation. We summarized the mass selection process in a cyclotron.

A given ion, with a mass $M_1$ and charge $Q_1$, has to fulfil the synchronism condition to be properly accelerated :

$$f_{rev} = \frac{Q_1}{2\pi M_1} \frac{B(r)}{\gamma(r)} = f_{RF}/H \quad (I.1)$$

Where $f_{rev}$ and $f_{RF}$ are the revolution and the RF frequencies, B(r) the magnetic field and γ the relativistic factor.

For another ion, with a different mass to charge ratio ($M_2/Q_2$), the synchronism condition is not fulfilled, the relative phase $\phi_2$ with the RF system evolves up to $\phi_2$=90°, and then these ions are not accelerated anymore (figure 4).

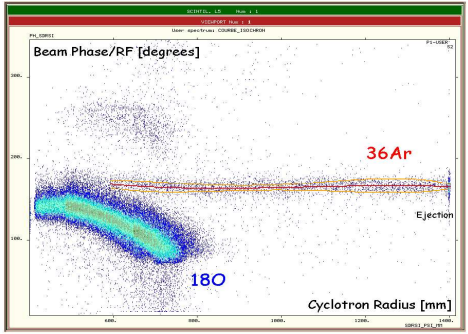

*Fig 4 : Phase measurement as a function of the radius in CIME using a 300 µm silicon detector. The $^{18}O^{4+}$ beam is lost, while the $^{36}Ar^{8+}$ ions continue toward the cyclotron extraction. Test realised on harmonic 3.*

The condition to eliminate an undesired ion corresponds at first order to:

$$\left[\frac{M_2}{Q_2} - \frac{M_1}{Q_1}\right]/\frac{M_1}{Q_1} > \frac{1}{2\pi H N_{turn}} \quad (I.2)$$

where $N_{turn}$ is the number of turns. Hence the mass resolution of a cyclotron is defined as $R = 1/2\pi H N_{turn}$.

Depending on the harmonic, the mass resolution of CIME can reach $1.6 \cdot 10^{-4}$. The mass selection obtained experimentally, whatever the harmonic, is close to the one expected theoretically (fig 3). Besides the cyclotron ensure a very clean separation, since the ion contaminants with mass deviation greater than $5.10^{-4}$ are strongly suppressed : the transmission for these ions are smaller than $10^{-6}$.

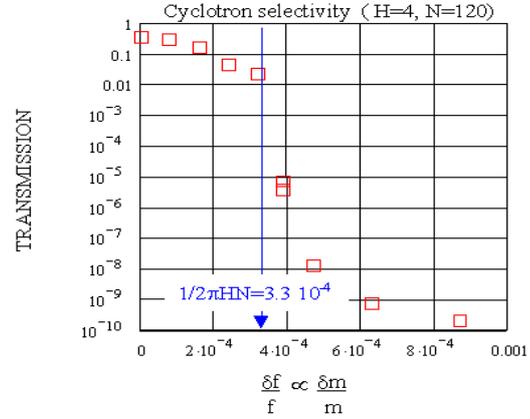

*fig.4 : Transmission of CIME cyclotron as a function of the mass over charge ratio deviation with the Harmonics 4.*

This study demonstrate that the nominal tuning, associated with the selectivity of the ECR ion source allow a total purification, provided that the charged state of radioactive ions are properly chosen. In the context of SPIRAL2, heavier ions (A>100) produced by non-chemically-selective sources could be accelerated with CIME. A much better mass resolution will be required to purify the beams. The possibility to achieve a resolution around $5.10^{-5}$ would be studied very soon.

## Tuning method of SPIRAL

The initial tuning of CIME is realised with a stable isotopic beam delivered by the ion source. The intensity available with this stable element allows the tuning of each accelerator section with classical diagnostics.

Strictly speaking, the only way to reproduce the same trajectories with the radioactive beam as the ones of the stable beam used for the tuning is to scale all the electric fields and magnetic fields with the factor c :

$$c = \frac{M_1}{Q_1} \frac{Q_2}{M_2} \quad (I.3)$$

This corrective factor applied on all the fields implies the conservation of the Newton-Lorentz equation.

Practically, the modification of all the equipments is not necessary especially if the mass over charge ratio of the two ions are very close.

Then, considering equation (I.1), there is two simplified strategies for the cyclotron to shift from one ion (M1,Q1) to another (M2, Q2) : a magnetic field (B) or RF frequency ($f_{RF}$) shift. Depending on the strategy chosen, and due to evolution of γ with acceleration, the synchronism condition may not be strictly respected at any radius. Nevertheless, as the γ factor at the exit of CIME is low, and as the (Q/M) ratio of stable and of radioactive beams are close, both methods may be convenient, but most of time, the RF frequency ($f_{RF}$) shift of the cyclotron is preferred.

As soon as the shift is performed, a global optimization of the radioactive beam is realized with the help of a germanium detector in the low energy beam line, a silicon detector in the cyclotron and gas profiles detector in the line which connect to the experimental hall.

## ECR SOURCES DEVELOPMENT

The improvement of the intensity and the variety of ion beams available rely on the evolution of source technology [6].

As an example, the MIVOC method (Metal Ions from Volatile Compounds) developed at Jyväskylä has been used at GANIL, since 1999, for the production of iron and nickel beams. 50 µA of $^{58}Ni^{11+}$, to get a 5 µA $^{58}Ni^{26+}$ on SISSI production target, have been produced several times.

Developments with natural Calcium, used under the metallic form, have enabled to get a relatively high intensity with $^{48}Ca^{10+}$ (enrichment 56%) : in June 2001, an average beam intensity of 15 µA of $^{48}Ca^{10+}$ was obtained during three weeks, with a good stability and a rather low consumption ($\leq 0.1$ mg/h).

A high intensity beam of $^{76}Ge^{10+}$ has bee obtained using a chemical reaction with $SF_6$ : it has been discovered that it was possible to recover germanium condensed on the plasma chamber. 35µA of $^{76}Ge^{10+}$ have been extracted from the source .

## OPERATION WITH INTENSE BEAMS

Longitudinal space charge effects in CSS2 represent a strong limitation for high intensity beam [4,5]. The increase of the energy spread of the bunches induces a turn overlap, which increase the beam losses at the cyclotron extraction. Thus, an increase in the dee voltage, operated generally at 170 kV (@13.45 MHz), was required. A detailed study of the RF cavities confirmed that voltages as high as 250 kV could be reached.

A first beam test was realized in December 2001, with a 95 MeV/A Argon beam, and a RF voltage of 200 kV. The overall transmission of the cyclotron was equal to 97%, and the ejected beam intensity equal to 26 µA, corresponding to a thermal power of 5 kW.

In the years 2001-2003, several improvements were made to facilitate the operation (tuning and supervision) of intense beams for use with SISSI and SPIRAL targets. These improvements concern the security of equipments, the beam stability, and the instrumentation needed to measure the beam characteristics. In addition, the project was extended to the possibility to send a high intensity beam directly to the experimental areas (LISE /LISE 2000 spectrometers). The routine operation of this high intensity began at the end of 2001, and was pursued these last two years, with for instance the acceleration of $^{13}C^{6+}$, with 3 kW beam power, to use with the SPIRAL target-source (production of $^8$He and post-acceleration with CIME).

Table 2: Example of stable beam available

| Beam | Energy (MeV/A) | Available intensity (CSS2 exit) |
|---|---|---|
| $^{13}C^{3/6}$ | 75 | 18 µA- 1.9 $10^{13}$ pps |
| $^{36}Ar^{10/18}$ | 95 | 26 µA- 9 $10^{12}$ pps |
| $^{36}S^{10/16}$ | 77 | 6,3 µA - 2.5 $10^{12}$ pps |
| $^{58}Ni^{11/26}$ | 74 | 5 µA – 1.2 $10^{12}$ pps |

To get better beam stability which is of crucial importance to minimize the beam losses, a deep modification of the high voltage platform dedicated to the injection in C01 cyclotron has been undertaken and completed in July 2004. A pre-selection stage has been added to decrease the total current in the acceleration tube. The beam is now extracted at 25 kVolts from the source mass selected and sent through the acceleration tube (<100kV). The first test show a remarkable stability of the beam, and since the total voltage of the platform can be increased. Besides we can inject in the cyclotrons chain lower charge states which result to a global improvement of the beam intensity.

## SPIRAL II PROJECT

Compared to SPIRAL I, SPIRAL II project aims to produce heavier (80<A<150), higher intensity secondary beams, by the ISOL method. A Linac driver will deliver a 5 mA, 40 MeV deuteron beam [7]. This beam will bombard a converter, producing neutrons which will induce up to $10^{14}$ fissions/s in an uranium target. The radioactive beams will be extracted from a source in a mono-charge state. A separator will allow the use of one beam at low energy, while a second will be available for charge breeding and post-acceleration with CIME.

## CONCLUSION

The main part of the activity of the facility concerns nuclear physics studies. GANIL-Spiral is producing routinely radioactive beams, with the two complementary techniques: in-flight projectile fragmentation and the so-called ISOL method. Numerous modifications of the facility have been accomplished to improve the intensity and the availability of the different ion beams. Let us cite:

- A new high voltage platform for the C01 cyclotron injector (2004)

- A new rotating target for in flight fragmentation production (available in 2005)

- An upgrade of the LISE spectrometer (2001-2002)

- A new beam line for material irradiation and radiobiology (2003)
- A new Beam line for low energy radioactive beams (2004)

In the near future, SPIRAL II project will require important R&D effort of the GANIL staff.